\documentstyle[11pt,epsfig,color,psfig,fleqn]{article}
\begin{document}
%---------------------------------------------------
 
%\begin{minipage}{16cm}
\title{ Evidence of $\sigma$ particle  
        in $J/\psi \to \omega \pi \pi$ \thanks{Data     
analyzed were taken prior to the participation      
of U.S. members and Japanese members of BES Collaboration. } }

%\begin{document}
%\date{}
\maketitle
\vskip -0.5cm

\vspace{0.3cm}
\begin{center}
(BES  Collaboration)
\end{center}
\vspace{0.3cm}
\par
\noindent
BAI Jing-Zhi,             BAN Yong$^{5}$,       BIAN Jian-Guo,  
CAI Xiao,                 CHANG Jin-Fan,        CHEN Hai-Xuan,  
CHEN He-Sheng,            CHEN Hong-Fang$^{1}$, CHEN Jiang-Chuan,     
CHEN Jin,                 CHEN Jun$^{11}$,      CHEN Ma-Li,
CHEN   Yuan-Bo,           CHI Shao-Peng,        CHU Yuan-Ping,
CUI Xiang-Zong,           DAI Hong-Liang ,      DAI You-Shan$^3$,     
DENG Zi-Yan,                                    DU Shu-Xian,
DU Zhi-Zhen,              FANG Jian,            FANG Shuang-Shi,
FU Cheng-Dong,            FU Hong-Yu,           FU Li-Ping$^{11}$,
GAO Cui-Shan,             GAO Mei-Li,           GAO Yuan-Ning$^{14}$, 
GONG Ming-Yu,             GONG Wen-Xuan,        GU Shu-Di,            
GUO Ya-Nan,               GUO Yi-Qing,          HAN Shi-Wen,          
HE Ju,                    HE Kang-Lin,          HE Mao$^{2}$,         
HE Xiang,                 HENG Yue-Kun,         HU Hai-Ming,
HU Tao,                   HUANG  Guang-Shun,     HUANG Liang$^{11}$,
HUANG Xiu-Ping,           JIA Qing-Ying$^{5}$,  JI Xiao-Bin,              
JIANG Chun-Hua,           JIANG Xiao-Shan,      JIN Da-Peng,              
JIN Shan,                 JIN Yan,              LAI Yuan-Fen,         
LI Fei,                   LI Gang,              LI Hui-Hong,        
LI Jia-Cai,               LI Jin,               LI Qiu-Ju,            
LI Ren-Ying,              LI Ru-Bo,             LI Shu-Min,       
LI Wei,                   LI Wei-Guo,           LI Xiao-Ling$^{12}$, 
LI Xue-Qian$^{8}$,        LI Xue-Song$^{14}$,   LIANG Yong-Fei$^{13}$,  
LIAO Hong-Bo$^{6}$,       LIU Chun-Xiu,         LIU Fang$^{1}$,
LIU Feng$^{6}$,           LIU Huai-Min,         LIU Jian-Bei,
LIU Jue-Ping$^{10}$,      LIU Rong-Guang,       LIU Yan,
LIU Zhen-An,              LIU Zhong-Xiu,        LU Gong-Ru$^{9}$,         
LU Feng,                  LU Jun-Guang,         LUO Cheng-Lin$^{16}$,     
LUO Xiao-Lan,             MA Feng-Cai$^{12}$,   MA Ji-Mao,  
Ma Lian-Liang$^{2}$,      MA Xiao-Yan,          MAO Ze-Pu,                
MENG Xiang-Cheng,         MO Xiao-Hu,           NIE Jing,                 
NIE Zhen-Dong,            PENG Hai-Ping$^{1}$,  QI Na-Ding,               
QIAN Cheng-De$^{4}$,      QIN Hu$^{16}$,        QIU Jin-Fa, 
REN Zhen-Yu,              RONG Gang,            Ruan Tu-Nan$^{1}$,
SHAN Lian-You,       
SHANG Lei,                SHEN Ding-Li,         SHEN Xiao-Yan,            
SHENG Hua-Yi,             SHI Feng,             SHI Xin$^{5}$,
SONG Li-Wen,              SUN Han-Sheng,        SUN Sheng-Sen$^{1}$,  
SUN Yong-Zhao,            SUN Zhi-Jia,          TANG Xiao,            
TAO Ning$^{1}$,           TIAN Yu-Run$^{14}$,   TONG Guo-Liang,           
WANG Da-Yong,             WANG Jin-Zhu,         WANG Lan,                 
WANG   Ling-Shu,          WANG Man,             WANG Meng,                
WANG Pei-Liang,           WANG Ping,            WANG Shu-Zhi,
WANG Wen-Feng,            WANG Yi-Fang,         WANG Zhe,
WANG Zheng,               WANG Zheng,           WANG Zhi-Yong,
WEI Cheng-Lin,            WU Ning,              WU Yuan-Ming,         
XIA Xiao-Mi,              XIE Xiao-Xi,          XIN Bo$^{12}$,        
XU Guo-Fa,                XU Hao,               XU Ye,
XUE Sheng-Tian,           YAN Mu-Lin$^{1}$,     YAN Wen-Biao,       
YANG Fan$^{8}$            YANG Hong-Xun$^{14}$, YANG Jie$^{1}$,     
YANG Sheng-Dong,          YANG Yong-Xu$^{15}$,  YI Li-Hua$^{11}$,    
YI Zhi-Yong,              YE Mei,               YE Ming-Han$^{7}$,    
YE Yun-Xiu$^{1}$,         YU Chuan-Song,        YU Guo-Wei,               
                          YUAN Jian-Ming,       YUAN Ye,                  
YUE Qian,                 ZANG Shi-Lei,         ZENG Yun$^{11}$,          
ZHANG Bing-Xin,           ZHANG Bing-Yun,       ZHANG Chang-Chun,         
ZHANG Da-Hua,             ZHANG Hong-Yu,        ZHANG Jia-Wen,            
ZHANG Jian,               ZHANG Jian-Yong,      ZHANG Jun-Mei$^{9}$,          
ZHANG Liang-Sheng,        ZHANG Qin-Jian,       ZHANG   Shao-Qiang, 
ZHANG Xiao-Mei,           ZHANG Xue-Yao$^{2}$,  ZHANG Yi-Yun$^{13}$,  
ZHANG Yong-Jun$^{5}$,     ZHANG Yue-Yuan,       ZHANG Zhi-Qing$^{9}$,
ZHANG Zi-Ping$^{1}$,      ZHAO Di-Xin,          ZHAO Jian-Bing,   
ZHAO Jing-Wei,            ZHAO Ping-Ping,       ZHAO Wei-Ren,  
ZHAO Xiao-Jian,           ZHAO Yu-Bin,             
ZHENG Han-Qing$^{5}$,     ZHENG Jian-Ping,      ZHENG Lin-Sheng,      
ZHENG Zhi-Peng,           ZHONG Xue-Chu,        ZHOU Bao-Qing,        
ZHOU Gao-Ming,            ZHOU Li,              ZHOU Neng-Feng,       
ZHU Ke-Jun,               ZHU Qi-Ming,          ZHU Ying-Chun,                    
ZHU Yong-Sheng,           ZHU Yu-Can,           ZHU Zi-An,
ZHUANG Bao-An,            

\vspace{1cm}
       Institute of High Energy Physics, Beijing 100039
\par
$^{1}$ University of Science and Technology of China, Hefei 230026
\par
$^{2}$ Shandong University, Jinan 250100
\par
$^{3}$ Zhejiang University, Hangzhou 310028
\par
$^{4}$ Shanghai Jiaotong University, Shanghai 200030
\par
$^{5}$ Peking University, Beijing 100871
\par
$^{6}$ Huazhong Normal University, Wuhan 430079
\par
$^{7}$ China Center for Advanced Science and Technology(CCAST), Beijing 100080
\par
$^{8}$ Nankai University, Tianjin 300071
\par
$^{9}$ Henan Normal University, Xinxiang 453002
\par
$^{10}$ Wuhan University, Wuhan 430072
\par
$^{11}$ Hunan University, Changsha 410082
\par
$^{12}$ Liaoning University, Shenyang 110036
\par
$^{13}$ Sichuan University, Chengdu 610064
\par
$^{14}$ Tsinghua University, Beijing 100084
\par
$^{15}$ Guangxi Normal University, Guilin 541004
\par
$^{16}$ Nanjing Normal University, Nanjing 210097

\normalsize 

\begin{abstract}

Based on a sample of $7.8 \times 10^6$ BESI $J/\psi$  
events, the decay of $ J/\psi \to \omega \pi^+ \pi^-$
is studied.  A low mass enhancement in the 
$\pi^+ \pi^-$ invariant mass spectrum 
recoiling against $\omega$ particle is clearly
seen which does not
come from the phase space effect and the background.
According to PWA analysis, this low mass enhancement
is a broad $0^{++}$ resonance, the $\sigma$ particle. 
If a Breit-Winger
function of constant width is used to fit 
the $\sigma$ signal,
its mass and width are $384 \pm 66$ MeV and  
$458 \pm 100$ MeV respectively, which correspond
to the pole position at 
(434 $\pm$ 78 ) - $i$ (202 $\pm$ 43) MeV. \\


\end{abstract}
%\clearpage

%\vskip 0.3in

%-------------------------------------------------------

\twocolumn

The evidence of  $\sigma$-particle is found in $\pi \pi$ $s$-wave.
The analysis of $\pi \pi$ phase shift obtained from CERN-Munich
experiment in 1974 found that the $I=0$ $\pi \pi$ S-wave phase
shift $\delta^0_0$ up to $m_{\pi \pi} = 1300$ MeV turned out to be
only 270$^{\circ}$ \cite{1}. After subtracting a rapid
contribution of the resonance $f_0(980)$ (180$^{\circ}$), the
remained phase  shift is  90$^{\circ}$. Thus, most analyses made
on the $\sigma$ particle have yielded conclusions against the
existence of $\sigma$. As a result, the light $\sigma$-particle
had been disappeared from the list of PDG since the 1976
edition\cite{3}. However, more and more evidences from both
experimental and theoretical analysis have been accumulated since
then in supporting the existence of the $\sigma$ particle. As a
result the $\sigma$ particle has reappeared in the recent editions
of the PDG. In the $p p$-central collision experiment, a huge
event concentration in $I=0$ S-wave $\pi \pi$-channel was seen in
the region of $m_{\pi \pi}$ around 500 $\sim$ 600 MeV\cite{4}.
This huge event concentration is too large to be explained as a
simple "background" and it strongly suggests the existence of
$\sigma$\cite{5}. 
\\

Various recent analysis on $\pi\pi$ phase shift data showed strong
evidence for the existence of the $\sigma$ particle, using
Breit--Wigner parametrization form\cite{7}, or other
parametrizations\cite{Tornq,61,71}. 
Applying analyticity and single-channel
unitarity, H.~Q.~Zheng proves that the $\sigma$ resonance is
necessary for chiral symmetry to explain the $\pi\pi$ scattering
process\cite{81}. A recent analysis based on Chiral Perturbation
Theory and the Roy equation has been made by Colangelo, Gasser and
Leutwyler, with the result of $M_{\sigma} = (490 \pm 30) - i (295
\pm 20)$ MeV\cite{8}.
\\

A broad low mass enhancement in  $\pi^+ \pi^-$
invariant mass spectrum in 
$J/\psi \to \omega \pi^+ \pi^-$ is observed by 
DM2\cite{9}, MARKIII\cite{10} and BES\cite{bes}. 
It is also observed in $D$ decay \cite{102} and 
$\Upsilon$  decay.
In this work, we study the structure of the
broad low mass enhancement in the $\pi^+ \pi^-$ 
invariant mass spectrum in $J/\psi \to \omega \pi^+ \pi^-$.
 \\

A sample of  $7.8 \times 10^6$ BESI 
$J/\psi$ events, which were accumulated by Beijing
Spectrometer\cite{11}, is used in the analysis.
Candidate tracks are required to have a good 
track fit with vertex position within the interaction 
region of 2 cm in $\sqrt{x^2+y^2}$ and 20 cm in $Z$
(beam direction), and in the polar angle region
of $|cos\theta|<0.8$. For neutral tracks,
it is required that the deposit energy of each
neutral track in Barrel Shower Counter(BSC) 
is greater than 50 MeV, the hitted showers 
start in the first six radiation length,
and the angle between
the direction from the event vertex 
to the position at the first layer of BSC and 
the developing direction of the cluster 
is less than $30^\circ$. 
Every candidate event  is then required to have
at least two neutral tracks, four charged tracks
with zero net charge. Time of Flight counter(TOF)
information is used for particle identification.
All surviving events are submitted to the
$\gamma \gamma  \pi^+ \pi^- \pi^+ \pi^-$ $4C$ kinematic 
fit and $\pi^0 \pi^+ \pi^- \pi^+ \pi^-$ $5C$ kinematic fit.
In the final data selection, 
$\chi^2_{4C} (J/\psi \to 2 \gamma \pi^+ \pi^- \pi^+ \pi^-) < 30$
and
$\chi^2_{5C} (J/\psi \to \pi^0 \pi^+ \pi^- \pi^+ \pi^-) <50$
are imposed.
The difference between the invariant
mass of two $\gamma$ and rest mass of $\pi^0$ is
less than 100 MeV. 
There are four combinations of
$\pi^0 \pi^+ \pi^-$. The one with the invariant mass 
closest to the rest mass of $\omega$ meson is regarded 
as the right combination.
\\

\begin{figure}[htbp]
\begin{center}
{\mbox{\epsfig{file=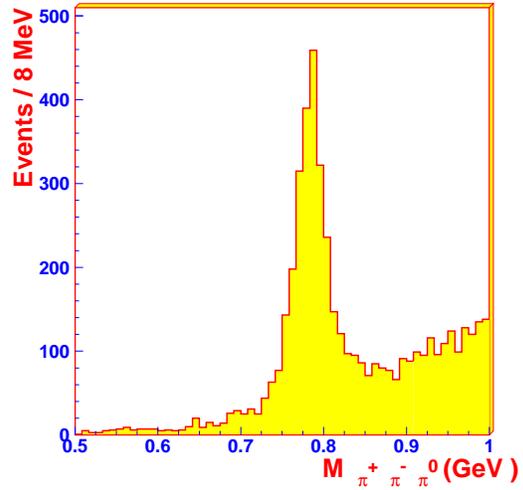,height=2.8in,width=2.8in}}}
\end{center}
\caption[]{  Invariant mass spectrum of $\pi^+ \pi^- \pi^0$. }
\label{w002a}
\end{figure}

\begin{figure}[htbp]
\begin{center}
{\mbox{\epsfig{file=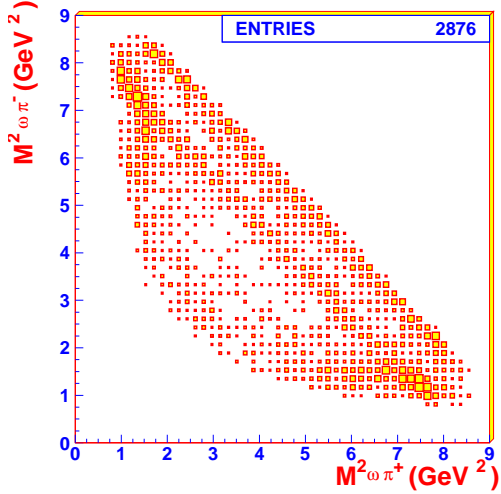,height=2.8in,width=2.8in}}}
\end{center}
\caption[]{ Dalitz plot of $M^2_{\omega \pi^-}$ vs 
	$M^2_{\omega \pi^+}$. The top
	slope band corresponds to the $\sigma$ particle,
	the second top slope band corresponds to the 
	$f_2(1270)$, the vertical band corresponds 
	$b_1(1235)^+$ and the horizontal band corresponds
	to $b_1(1235)^-$. }                                
\label{w002b} 
\end{figure}

\begin{figure}[htbp]
\begin{center}
{\mbox{\epsfig{file=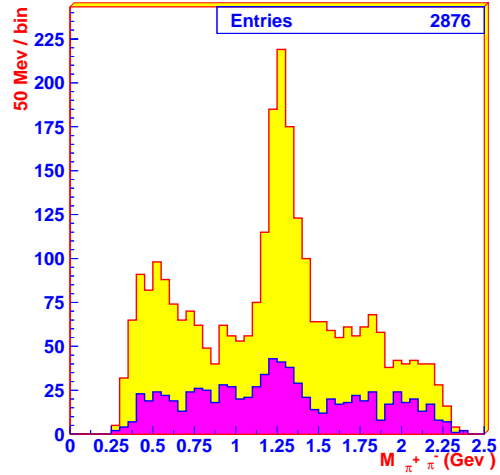,height=2.8in,width=2.8in}}}
\end{center}
\caption[]{ The invariant mass spectrum of $\pi^+ \pi^-$. 
	The heavy shaded region shows the $\omega$ side-band
	structure. }
\label{w03a}
\end{figure}

\begin{figure}[htbp] 
\begin{center}
{\mbox{\epsfig{file=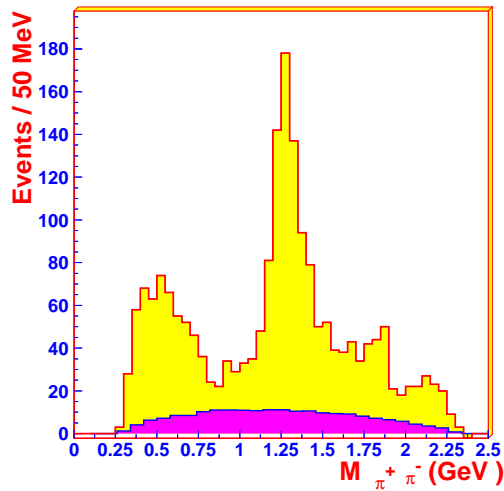,height=2.8in,width=2.8in}}}
\end{center}
\caption[]{ The invariant mass spectrum of $\pi^+ \pi^-$
        after $\omega$ side-band subtraction. The heavy shaded
        region shows the shape of phase space.
 	The broad low mass enhancement at about 500 MeV
	is the $\sigma$ particle, and the high peak
	at 1270 MeV is $f_2(1270)$. }  
\label{w03b} 
\end{figure}

Fig.\ref{w002a}  shows the $\pi^0 \pi^+ \pi^-$ invariant
mass spectrum after the above cuts, where $\omega$
signal can be clearly seen with a low background level.
The $\pi^+ \pi^-$ invariant mass spectrum, recoiling
against $\omega$ with 
$|M_{\pi^+\pi^-\pi^0}-M_{\omega}|<0.05$,  
is shown in Fig.\ref{w03a} and the corresponding 
Dalitz plot is shown in Fig.\ref{w002b}.
A broad low mass enhancement can be clearly seen
in $\pi^+\pi^-$ invariant mass spectrum, 
the corresponding
band can also be seen in the Dalitz plot,
 and the  shape of the low mass enhancement
is quite different from that of the phase
space(Fig.\ref{w03b}).
If the broad low mass enhancement came from 
the phase space effect, the corresponding events
would be uniformly scattered in the whole 
Dalitz plot region, so there would be no band 
corresponding to it. Therefore, 
this low mass enhancement does not originate
from phase space effect. 
We can also see that the
band is evenly distributed which is the signature of
a $0^{++}$ resonance. In figure \ref{w03a},
the heavy shaded histogram shows the
$\pi^+ \pi^-$ invariant mass which recoils against
$\omega$ side-band where no clear 
low  mass structures was found.
Fig.\ref{w03b} shows the $\pi^+ \pi^-$  
invariant mass spectrum after $\omega$ side-band
subtraction. The low mass enhancement can still
be clearly seen, which means that  it does not
come from those background channels which contain
no $\omega$ particle in its decay sequence. \\

We perform detailed Monte Carlo study and find 
that the background events from other known $J/\psi$
decay channels are very small. The background
events from other known $J/\psi$ decay channels 
can not produce an enhancement here. 
\\

According to the above study, we know that the low
mass enhancement in the $\pi\pi$ invariant mass
spectrum does not come from the phase space effect
and the background.  We perform Partial
Wave Analysis(PWA) on the $\pi^+ \pi^-$
invariant mass spectrum to study the structure of 
the low mass enhancement. 
Covariant helicity coupling amplitude method
is used\cite{14,15}.
The analysis method and the theoretical formula
used in our analysis can be found in
literature \cite{15}. $\sigma$, $f_2(1270)$,
$f_0(980)$, $f_0(1710)$ and $f_2(2300)$
are considered as contributions 
in the $\pi^+ \pi^-$ invariant mass spectrum,
$b_1(1235)$ is considered in the 
$\omega \pi$ invariant mass spectrum. 
We add an non-interference amplitude to fit
backgrounds from $J/\psi \to \rho 3 \pi$
directly. All other backgrounds are approximated
by a non-interference phase space background.  \\

Mass and width of $\sigma$ particle are somewhat
model dependent. In this paper, 
several Breit--Wigner functions are tried 
to fit $\sigma$ particle\cite{pdg,16,71}

\begin{equation}
\left \lbrace
\begin{array}{l} 
BW_{\sigma} = \frac{1}{m_{\sigma}^2 - s - 
i m_{\sigma} \Gamma_{\sigma}} \\
\Gamma_{\sigma}{\rm~is~a~constant}
\end{array}
\right .,
\label{01}  
\end{equation}

\begin{equation}
\left \lbrace
\begin{array}{l}  
BW_{\sigma} = \frac{1}{m_{\sigma}^2 - s - i \sqrt{s}
\Gamma_{\sigma}(s)}  \\
\Gamma_{\sigma}(s) 
=\frac{g_{\sigma}^2 \sqrt{\frac{s}{4} - m_{\pi}^2 } }
{8 \pi s}
\end{array} 
\right . ,
\label{02}
\end{equation}

\begin{equation}   
\left \lbrace 
\begin{array}{l}  
BW_{\sigma} = \frac{1}{m_{\sigma}^2 - s - i \sqrt{s}
\Gamma_{\sigma}(s)}  \\ 
\Gamma_{\sigma}(s)      
= \alpha \sqrt{\frac{s}{4} - m_{\pi}^2 } 
\end{array}
\right . ,
\label{03}  
\end{equation}

\begin{equation}
\left \lbrace
\begin{array}{l}
BW_{\sigma} = \frac{1}{m_{\sigma}^2 - s - i m_{\sigma}
( \Gamma_1 (s)+ \Gamma_2 (s) )},  \\
\\
\Gamma_1 (s) = G_1 
\frac{\sqrt{1 - 4 m_{\pi}^2/s}}
{\sqrt{1 - 4 m_{\pi}^2/m_{\sigma}^2}} \cdot \\
~~~~~~~~~~~ \cdot
\frac{s-m_{\pi}^2/2}{m_{\sigma}^2-m_{\pi}^2/2}
e^{-(s-m_{\sigma}^2)/4 \beta^2},  \\
\\
\Gamma_2 (s) = G_2
\frac{\sqrt{1 - 16 m_{\pi}^2/s}}
{\sqrt{1 - 16 m_{\pi}^2/m_{\sigma}^2}} 
\frac{1 + e^{\Lambda (s_0 - m_{\sigma}^2)}}
{1 + e^{\Lambda (s_0 - s)}}.
\end{array}
\right .
\label{04}  
\end{equation}
where parameter $G_1$ = 1.378 GeV, $\beta$ = 0.7 GeV
$G_2$ = 0.036 GeV and $\Lambda$ = 3.5 GeV$^{-1}$ \cite{71}.
The width of $\sigma$ particle given in this paper
is the width at its mass, i.e., 
$\Gamma_{\sigma}(m_{\sigma})$.
We first use a $0^{++}$ to fit the first peak, 
then we test its statistical significance and spin-parity.
It is found that $\sigma$ particle has about
18 $\sigma$ statistical significance in this channel.
If we change its
spin-parity from $0^{++}$ to $2^{++}$ (or $4^{++}$), 
the log likelihood will become worse by 
about 29 (or 37), which is a 6.6 (or 7.7) $\sigma$
effects.  So, its  spin-parity should be $0^{++}$.
Mass and width of $\sigma$ particle are determined 
through mass and width scan. Our final
results on mass, width and pole of $\sigma$ particle
are listed in Table 1.
If we use eq. (\ref{01}) to fit sigma, its
branching ratio is
$
BR(J/\psi \to \omega \sigma \to \omega \pi^+ \pi^- )
= (11.7 \pm 1.7 \pm 5.6 ) \times 10^{-4},
$
where the  first error is  statistical 
and the second is systematic. The systematic
error comes from the different parametrization
of the Breit-Wigner formula(39\%), 
the uncertainty of the detection
efficiency(20\%) and the uncertainty of the 
$J/\psi$  total number(20\%).   \\

{\footnotesize
\begin{table}[htp]  
\begin{center} 
\doublerulesep 0pt     
\renewcommand\arraystretch{1.5}  
\begin{tabular}{|l|l|l|l|} 
\hline 
\hline  
\hline    

BW Function & Mass (MeV)  & width (MeV)  & Pole (MeV) \\
\hline

Eq.(\ref{01}) & 384 $\pm$ 66  &  458 $\pm$ 100 &
(434 $\pm$ 78 ) \\
&&& ~- $i$ (202 $\pm$ 43) \\
\hline

Eq.(\ref{02}) & 442 $\pm$ 28  &  346 $\pm$ 86 &
(470 $\pm$ 39 ) \\
&&& ~- $i$ (164 $\pm$ 38) \\
\hline

Eq.(\ref{03}) \cite{16} & 559 $\pm$ 52  &  566 $\pm$ 136 &
(432 $\pm$ 54 ) \\
&&& ~- $i$ (179 $\pm$ 28) \\
\hline

Eq.(\ref{04}) \cite{71} & ----  &  ----  &
(384 $\pm$ 15 )\\
&&& ~ - $i$ (285 $\pm$ 30) \\
\hline

\hline   
\hline   
\hline 
\end {tabular} 
\caption{Masses, widths and poles of $\sigma$ particle }     
\end{center}  
\end{table}   
}

\begin{figure}[htbp]
\begin{center}
{\mbox{\epsfig{file=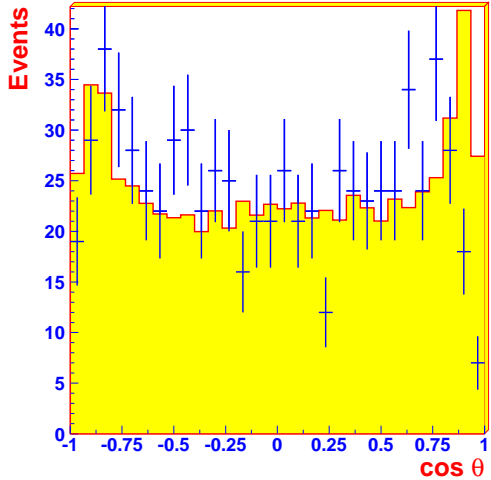,height=2.8in,width=2.8in}}}
\end{center}
\caption[]{  Final fit to the angular distributions for events
	in $\sigma$ mass region ($M_{\pi^+\pi^-}<0.8 $ GeV).
	Error bar is the data and histogram 
	is our final fit using eq. (1). 
	$\theta$ is the  polar angle of  
	$\pi^+$ in $\pi^+ \pi^-$ rest frame.}
\label{w16a}  
\end{figure}

\begin{figure}[htbp] 
\begin{center}
{\mbox{\epsfig{file=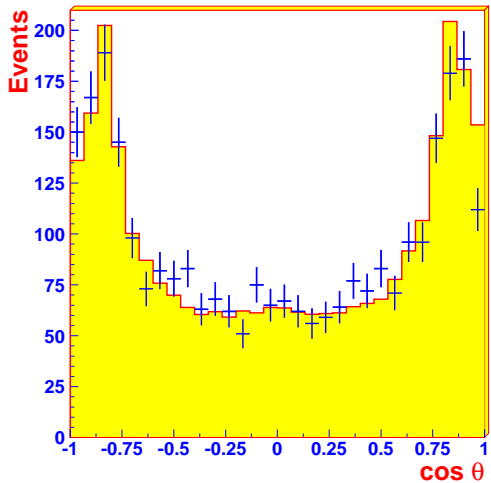,height=2.8in,width=2.8in}}}
\end{center}
\caption[]{  Final fit to the angular distributions for events  
	in the whole mass region. 
	Error bar is the data and histogram 
	is our final fit using eq. (1). 
	$\theta$ is the  polar angle of
	$\pi^+$ in $\pi^+ \pi^-$ rest frame.}
\label{w16b}
\end{figure}

\begin{figure}[htbp]
\begin{center}
{\mbox{\epsfig{file=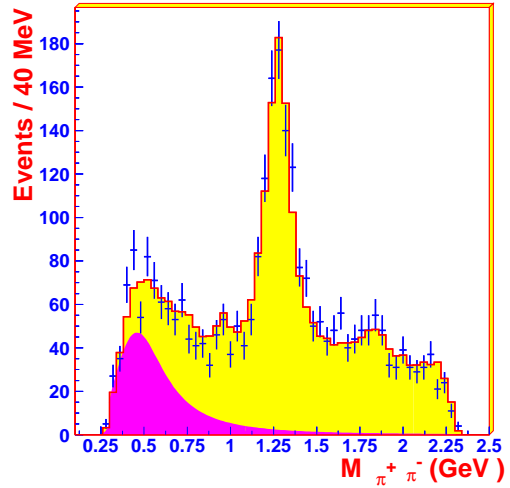,height=2.8in,width=2.8in}}}
\end{center}
\caption[]{ Final fit on the $\pi^+ \pi^-$ invariant
	mass spectrum recoil against $\omega$ particle.
	The error bar is the real data, 
	the shaded histogram is the  global fit and the
	heavy shaded curve region is the contribution 
	from $\sigma$-particle. }
\label{w17}
\end{figure}

The final fit on the  angular $\theta$ distributions 
for events in $\sigma$ mass region
($M_{\pi^+\pi^-}<0.8 $ GeV ) is shown in 
Fig.\ref{w16a}, where $\theta$ is the polar 
angle of $\pi^+$ in the  $\pi^+ \pi^-$ rest frame. 
In this channel, besides $\sigma$ particle,
$f_2(1270)$, $f_0(980)$, $b_1(1235)^+ \pi^-$
and $b_1(1235)^- \pi^+$ are added into the final
fit. For the whole mass region, Fig.\ref{w16b} 
shows the fit of the $\theta$ angular distributions,
while Fig.\ref{w17} shows the fit of the $\pi^+ \pi^-$ 
invariant mass spectrum, where $\sigma$ contribution  
is indicated by the heavy shaded region.  \\

In summary, 
a broad low mass enhancement is found in the recoil
$\pi^+ \pi^-$ invariant mass spectrum against
$\omega$ particle in 
$J/\psi \to \omega \pi^+ \pi^-$, and the correponding
band can be clearly seen in the Dalitz plot
( Fig.\ref{w002b}). According to the detailed
background study, the low mass enhancement does
not come from the backgrounds of other $J/\psi$
decay channels. It also does not come from
the phase space effects, and
is considered to be the $\sigma$ particle.
the existence of the $\sigma$ particle
can  be clearly seen in both the $\pi^+ \pi^-$
invariant mass spectrum and the Dalitz plot. 
According to PWA analysis, $\sigma$ particle
is highly needed in the final fit and its spin-parity
is $0^{++}$. The mass and width are somewhat 
model-dependent, but its pole position is
much less dependent on theory model. If a Breit-Winger
function with constant width is used to fit 
$\sigma$ particle,
its mass and width are $384 \pm 66$ MeV and  
$458 \pm 100$ MeV respectively, which correspond
to the pole position of
(434 $\pm$ 78 ) - $i$ (202 $\pm$ 43) MeV.
Its branching ratio is
$
BR(J/\psi \to \omega \sigma \to \omega \pi^+ \pi^- )
= (11.7 \pm 1.7 \pm 5.6 ) \times 10^{-4}.
$
\\

\lbrack {\bf ACKNOWLEDGEMENT} \rbrack:
We would like to thank Prof. S. Ishida,
Dr. M. Ishida, Dr. T. Komada, Prof. K. Takamatsu, Prof. T.Tsuru,
and Prof. K. Ukai  for usefull discussions
on $\sigma$-particles.
The BES collaboration thanks the staff of BEPC for
their hard efforts. This work is supported in part 
by the National Natural Science Foundation
of China under contracts Nos. 19991480, ~10225524,
10225525, the Chinese Academy of Sciences under 
contract No. KJ 95T-03, the 100 Talents Program of 
CAS under Contract Nos. U-11, U-24, U-25, and the 
Knowledge Innovation Project of CAS under Contract 
Nos. U-602, U-34 (IHEP); and by the National Natural 
Science Foundation of China under Contract 
No.10175060 (USTC).


\vspace {-0.2in}

\begin{thebibliography}{99}



\bibitem{1} G.Grayer $et~ al$, Nucl. Phys. {\bf B75}, 189
        (1974); B.Hyams $et~ al$, Nucl. Phys. {\bf B64},
        134(1973).

\bibitem{3} Review of Particle Properties,
        $Rev. ~ Mod.~ Phys.$ {\bf 48} S114 (1976).

\bibitem{4}     D.Alde $et~ al$, $Phys.~ Lett.$
        {\bf B397}, 350(1997).

\bibitem{5}    T.~Ishida $et~al$, in proceedings of Int.Conf.
        Hadron'95, p.451, Manchester UK, July  1995(World Scientific).


\bibitem{7}    S.~Ishida $et~al.$,  Prog. Theor. Phys.
        {\bf 95}, (1996) 745;
    S.~Ishida $et~al.$,  Prog. Theor. Phys.
        {\bf 98}, (1997) 1005
\bibitem{Tornq}N.~A.~T\"ornqvist, Z.~Phys.~{\bf C68}(1995)647.
\bibitem{61} E.~Van~Beveren, $et~ al.$,
    Eur.~Phys.~J. {\bf C22}, 493 (2001).

\bibitem{71} D.V. Bugg, A.V. Sarantsev, B.S. Zou,
    Nucl.Phys.B471:59-89,1996.

\bibitem{81}  Zhiguang Xiao, Hanqing Zheng,
    Nucl.Phys.{\bf A 695} (2001) 273-294.

\bibitem{8} G. Colangelo, J.Gasser and H. Leutwyler,
    Nucl. Phys. {\bf B603} (2001) 125.


\bibitem{9}    J.-E.Augustin $et ~al$, Proc. of the
        XXIth rencontre de Moriond, Les Arcs,
        9 -- 16  March, 1986, P.421.

\bibitem{10}    U.Mallik, Talk given at the SLAC Summer Institute
        on Particle Physics, Stanford 1986,
        SLAC-PUB-4238(1987).

\bibitem{bes} Ning Wu (BES collaboration), {\it BES R measurements
        and $J/\psi$ Decays}, Proceedings of the XXXVIth
        Rencontres de Moriond, Les Arcs, France,
        March 17 -- 24, 2001, Ed. J. Tran Thanh Van.
        2001 QCD and High Energy Hadronic Interactions,
        p. 3-6; hep-ex/0104050.


\bibitem{102}  E.M. Aitala et al. (E791 Collaboration),
    Phys.Rev.Lett. 86 (2001) 770-774.

\bibitem{11}  J.Z.Bai, {\it et al.} Nucl. Instrum. Methods Phys. Res.
     A {\rm 344 }, 319 (1994).

\bibitem{pdg} Particle Data Group, Euro. Phys. J. C15 (2000) 1.

\bibitem{12} J.C. Chen {\it et al.}, Phys.Rev.{\bf D62},
    034003 (2000)

\bibitem{13} J.Z.Bai {\it et al} " First evidence
    of  $\kappa$ particle in
    $J/\psi \to \bar{K}^*(892)^0 K^+ \pi^-$ "; hep-ex/0304001.

\bibitem{14} M.Jacob, G.C.Wick, Ann.Phys. (NY)7, 1959:404;
        S.U.Chung, Phys. Rev. D57, 1998:431-442;

\bibitem{15} Ning Wu and Tu-Nan Ruan, Commun. Theor. Phys.
        (Beijing, China) {\bf 35} (2001) 547;
        Ning Wu and Tu-Nan Ruan, Commun. Theor. Phys.
        (Beijing, China) {\bf 37} (2002) 309.

\bibitem{16} Hanqing Zheng, {\it How to parameterize a
    resonance with finite width }, Talk given at
    International Symposium on Hadron Spectroscopy,
    Chiral Symmetry and Relativistic Description of
    Bound Systems, Tokyo, Japan, 24-26 Feb 2003;
    hep-ph/0304173.



\end{thebibliography}
\end{document}